# ORIGINS OF THE HIGHLY IONIZED GAS ALONG THE LINE OF SIGHT TOWARDS HD 116852

ANDREW J. FOX[1], BLAIR D. SAVAGE[1], KENNETH R. SEMBACH[2], DIRK FABIAN[1], PHILIPP RICHTER[1], DAVID M. MEYER[3], JAMES LAUROESCH[3], AND J. CHRISTOPHER HOWK[4]

## ABSTRACT

We present Hubble Space Telescope Imaging Spectrograph (HST/STIS) and Far Ultraviolet Spectroscopic Explorer (FUSE) observations of high ion interstellar ultraviolet absorption along the sight line to HD 116852. At a distance of 4.8 kpc, HD 116852 is an O9 III star lying in the low Galactic halo, −1.3 kpc from the plane of the Galaxy in the direction $l = 304.9°$, $b = −16.1°$. The sight line passes underneath the Sagittarius-Carina and the Norma-Centaurus spiral arms. The STIS E140H grating observations provide high-resolution (FWHM ≈ 2.7 km s$^{-1}$) spectra of the resonance doublets of Si IV, C IV, and N V. These data are complemented by medium-resolution (FWHM ≈ 20 km s$^{-1}$) FUSE spectra of O VI. The integrated logarithmic column densities are log $N$(Si IV) = 13.60 ± 0.02, log $N$(C IV) = 14.08 ± 0.03, log $N$(N V) = 13.34 ± $^{0.05}_{0.06}$, and log $N$(O VI) = 14.28 ± 0.01. We find evidence for three distinct types of highly ionized gas present in the data. First, two narrow absorption components are resolved in the Si IV and C IV profiles, at approximate LSR velocities of −36 and −10 km s$^{-1}$. These narrow components appear to be produced in gas associated with the Norma and Sagittarius spiral arms, at approximate $z$-distances of −1.0 and −0.5 kpc, respectively. The temperature of the gas in these narrow components, as implied by their $b$-values, suggests that the gas is photoionized. The ratio of C IV to Si IV in these narrow components is low compared to the Galactic average. Second, we detect an intermediate-width component in C IV and Si IV, at 17 km s$^{-1}$, which we propose could arise at the conductive interface at the boundary between a low column density neutral or weakly ionized cloud and the surrounding hot medium. Finally, a broad collisionally ionized component of gas responsible for producing the smooth N V and O VI profiles is observed; such absorption is also present to a lesser degree in the profiles of Si IV and C IV. The broad O VI absorption is observed at a velocity displaced from the broad C IV component by almost 20 km s$^{-1}$, an amount large enough to suggest that the two ions may not co-exist in the same physical location. If these two ions do exist together, then the ratio $N$(C IV)/$N$(O VI) is too low to be consistent with turbulent mixing layer models, but could be explained by radiative cooling or conductive heating models. Combining our results with high resolution observations of four other sight lines from the literature, we find an average C IV component frequency of $1.0 \pm 0.25$ kpc$^{-1}$.

---

[1] Department of Astronomy, University of Wisconsin - Madison, 475 North Charter Street, Madison, WI 53706; fox@astro.wisc.edu, savage@astro.wisc.edu, richter@astro.wisc.edu
[2] Space Telescope Science Institute, 3700 San Martin Drive, Baltimore, MD 21218; sembach@stsci.edu
[3] Department of Physics and Astronomy, Northwestern University, 2145 Sheridan Road, Evanston, IL 60208; davemeyer@nwu.edu, jtl@elvis.astro.nwu.edu
[4] Department of Physics and Astronomy, Johns Hopkins University, 3400 North Charles Street, Baltimore, MD 21218; howk@pha.jhu.edu

*Subject headings*: Galaxy: halo – ISM: structure – stars: individual (HD 116852) – ultraviolet: ISM

## 1. INTRODUCTION

The warm and hot components of the interstellar medium can be investigated by analyzing absorption lines present in the spectra of background sources of ultraviolet (UV) radiation. The ions C IV, Si IV, and N V have previously been observed and studied in the Galactic disk and halo with the *International Ultraviolet Explorer* (IUE) and the *Goddard High Resolution Spectrograph* (GHRS) on-board the *Hubble Space Telescope* (HST). The GHRS was replaced in 1997 with the *Space Telescope Imaging Spectrograph* (STIS), an instrument capable of two-dimensional imaging as well as spectroscopy (Woodgate et al. 1998). None of these instruments were able to observe into the far-ultraviolet region (912 – 1150 Å), because they were not designed with the windowless detectors and specially coated optical components required for spectroscopy in this bandpass. In particular, the important O VI λλ1031.926, 1037.627 doublet remained unobservable with these instruments. Observations of O VI are significant to our understanding of the interstellar medium (ISM) because it is an effective tracer of collisionally ionized hot gas. Normal hot stars have a strong He II absorption edge, which limits the emergent flux of photons above 54 eV. Since the ionization potential for the production of O VI is 113.9 eV, it cannot be produced by photoionization from normal starlight. While pioneering studies of interstellar O VI toward bright stars were undertaken with the *Copernicus* satellite (Jenkins 1978a, b), the first instrument capable of observing the O VI doublet in the spectra of large numbers of faint sources came into service with the successful launch and commissioning of the *Far Ultraviolet Spectroscopic Explorer* (FUSE) in 1999 (Moos et al 2000).

In this paper we present high-resolution (FWHM ≈ 2.7 km s$^{-1}$) STIS observations of C IV, Si IV, and N V together with medium resolution (FWHM ≈ 20 km s$^{-1}$) FUSE observations of O VI along the 4.8 kpc sight line to the halo star HD 116852. The sight line is of interest because it passes through the Galactic disk, underneath the Sagittarius and Norma spiral arms, and out into the low halo. A previous study of the HD 116852 sight line (Sembach & Savage 1994, hereafter SS94) used medium resolution GHRS data (FWHM = 11 – 18 km s$^{-1}$) to investigate the structure and origin of the ISM in this direction. SS94 detected N V with a total column density of log $N$ = 13.48 ± 0.06, indicating the presence of hot gas ($T \sim 2 \times 10^5$ K) in the interstellar medium (ISM) in this direction. This suggests that the O VI λλ 1031.926 1037.627 doublet will be a useful diagnostic for further investigation of the hot ISM along this sight line. It is the goal of this paper to understand how the O VI distribution along the sight line relates to the high ion distribution discussed by SS94, and also to use the high resolution STIS data to refine our understanding of the C IV, Si IV, and N V distributions. Basic sight line information for HD 116852 is presented in Table 1.

We follow the procedures used in similar high-resolution studies of the HD 167756 (Savage, Sembach, & Cardelli 1994) and HD 177989 (Savage, Sembach, & Howk 2001) sight lines. These studies have been able to resolve the component structure of high ion absorption, providing new information about the nature and location of interstellar gas extending into the halo of the Galaxy.

This paper is organized in the following manner: §2 describes the observations and data processing, together with the methods used to calibrate the instrumental wavelength scales. In §3 a discussion of the sight line toward HD 116852 is presented. In §4 the main results of this paper are given: equivalent widths, total column densities, profiles of apparent column density as a function of velocity, component fits to the line structure, and component-to-component line ratios. Section 5 contains a discussion of the results, with regard to the gas kinematics and ionization at different locations along the sight line. In §6 we discuss the frequency of occurrence of highly ionized gas components in the ISM of the disk and low halo. Our results are summarized in §7.

## 2. OBSERVATIONS AND VELOCITY CALIBRATION

The observations that produced the spectra analyzed in this paper are described in Table 2. The STIS observations were made under Guest Observer programs GO8862, "Snapshot Survey of the Hot ISM" and GO8241, "Snapshot Survey of Interstellar Absorption Lines", whereas the FUSE observations were obtained as part of the FUSE science team O VI survey. In the following two sub-sections we discuss the data reduction and handling procedures for the two separate data sets.

### 2.1 STIS

The STIS observations were taken with HD 116852 in the 0.2″ x 0.2″ aperture, using the E140H high-resolution echelle grating to disperse the ultraviolet light onto the far-ultraviolet Multi-Anode Microchannel-Array (MAMA) detector. This grating affords a spectral resolution of $R \approx 110,000$, corresponding to a velocity FWHM of approximately 2.7 km s$^{-1}$. The STIS observations were treated with the CALSTIS (v2.3) processing pipeline to provide image linearization, dark-count subtraction, flat-field division, spectral order identification, and wavelength and photometric calibration. The resulting spectrum has been corrected using the scattered-light routine of Howk & Sembach (2000). For a description of the design and construction of STIS, see Woodgate et al. (1998). A summary of the STIS on-orbit performance is given by Kimble et al. (1998).

In order to calibrate the velocity scale to the LSR system a correction of $\Delta v_{LSR} = v_{LSR} - v_{Helio} = -6.4$ km s$^{-1}$ was applied, under the assumption that the solar neighborhood has a speed of 16.5 km s$^{-1}$ in the direction $l = 53°$, $b = 25°$ (Mihalas & Binney 1981). To check the accuracy of this wavelength scale, we measured the observed wavelength of the weak O I $\lambda$1355.598 line, and after converting to a velocity scale and applying the $\Delta v_{LSR}$ correction described above, it was found that $v_{LSR}$(O I) = 2.7 km s$^{-1}$. High-resolution Na I and Ca II spectroscopy along this sight line has been obtained by Sembach, Danks, & Savage (1993), who found a multi-component structure with the strongest Na I absorption at 2.7 km s$^{-1}$, and the strongest component of Ca II at 2.3 km s$^{-1}$. This suggests that the STIS velocity scale is accurate to within ±1 km s$^{-1}$, since we expect O I, Na I and Ca II to trace the same neutral component of the interstellar medium. Furthermore, we observed the Cl I $\lambda$1347.240 line, and using a Gaussian profile fitting procedure determined the line center to be at 1.9 km s$^{-1}$. This proved to be useful in calibrating the velocity of molecular hydrogen (H$_2$) absorption in the FUSE data, since Cl I and H$_2$ exist in the same interstellar regions (see §2.2).

2.2 FUSE

Twelve individual exposures of HD 116852 were obtained by FUSE on 27 May 2000, with the star centered in the LiF1 LWRS (30" x 30") aperture. We chose to analyze data from the LiF1A detector segment as it produced a higher signal-to-noise and higher resolution spectrum near the O VI line than the other FUSE channels covering that wavelength range. The raw histogram data were treated using the CALFUSE (v1.8.7) data reduction pipeline. The individual (LiF1A) exposures were co-added with no wavelength shift to produce the combined spectrum used in this paper. For a discussion of the spectroscopic capabilities and on-orbit performance of FUSE, we refer the reader to Moos et al. (2000) and Sahnow et al. (2000), respectively.

Calibrating the FUSE data to the LSR velocity scale involved a special correction procedure, since the wavelengths from the data reduction pipeline required a velocity correction of ~50 km s$^{-1}$ to bring them into the LSR reference frame. The approach taken to calibrate the FUSE spectrum in the region around the O VI λ1031.926 line was to measure the velocity of the molecular hydrogen (6-0) R(4) λ1032.356 line. A Gaussian component fitting procedure was used to model this line with two components. The stronger of these two components was centered at a measured velocity of 54.1 km s$^{-1}$. We were able to model the (6-0) P(3) λ1031.191 line with one Gaussian centered at 54.2 km s$^{-1}$, indicating the wavelength (and hence velocity) scale was consistent over the region near the O VI line of interest. We assumed that the strongest component of molecular hydrogen along the sight line to HD 116852 coincides exactly with the strongest component of neutral chlorine. This logic was based on our understanding of the chemistry of chlorine in the interstellar medium; neutral chlorine has an ionization potential of 12.97 eV (<13.54 eV, the Lyman limit), so that isolated interstellar Cl I atoms are photoionized by ambient photons and converted into Cl II. In the presence of H$_2$, Cl II undergoes a charge-exchange reaction that returns the ion to the Cl I state (Jura & York 1978). Consequently, H$_2$ suppresses the ionization of chlorine, and so H$_2$ and Cl I are expected to exist in the same interstellar regions. As discussed in §2.1, Cl I was measured in the STIS data to be strongest at v$_{LSR}$ = 1.9 km s$^{-1}$. Thus, a correction of 1.9 − 54.1 = −52.2 km s$^{-1}$ was required to bring the FUSE velocity scale into the LSR frame. Our whole calibration process is dependent on the coincidence of Cl I and H$_2$ absorption, which we believe to be reasonable because of the arguments given above. However, we still treat our FUSE velocity scale with some degree of caution, and estimate that a residual velocity offset of up to ±5 km s$^{-1}$ is possible. We note that the second component of the O VI doublet at 1037.627 Å was not useful in our investigation because of severe blending with C II$^*$ at 1037.018 Å and the H$_2$ (5–0) R(1) and P(1) lines at 1037.146 Å and 1038.156 Å.

For a line of sight with strong H$_2$ absorption, it is necessary to account for blending between the O VI λ1031.926 line and the deuterated hydrogen (HD) Lyman series (6–0) R(0) λ1031.912 line. To assess whether or not there is sufficient HD along this sight line to produce any significant blending, we searched for other (uncontaminated) HD lines in the LiF1A spectrum and found four "clean" profiles of HD absorption out of the ground state (J=0). No excited state (J=1 or above) transitions are evident in the spectrum. The rest wavelengths, oscillator strengths, equivalent widths and measured velocities of these lines are given in Table 3. The central velocities were estimated by fitting Gaussian components to the data. We have included a row for the HD (6–0) R(0) line, which is responsible for contaminating the O VI profile. The observed central velocities of the (5–0) R(0) and (4–0) R(0) lines are almost 10 km s$^{-1}$ smaller than the central velocities of the (7–0) R(0) and (8–0) R(0) lines. We interpret this difference as being caused by a wavelength-dependent velocity calibration error present in the data, since these components should occur at the same velocity. To account for these velocity shifts, we applied a correction so that all the HD profiles lined up with the H$_2$ 1032.356 Å line centered at v$_{LSR}$ = 1.9 km s$^{-1}$. The widths of the four observed profiles also varied slightly, due to the wavelength dependence of the instrumental line spread function. It was then necessary to combine these four HD profiles in such a way as to predict what the (6–0) R(0) profile would look like in the absence of O VI. By applying a curve of growth analysis to the four observed HD lines, we estimated an equivalent width of 15.6 mÅ for the (6-0) R(0) line. We modeled the contaminating HD line by combining the equivalent width estimate with the average instrumentally blurred profile width (FWHM = 29 km s$^{-1}$) measured for the other HD lines. This model profile was then divided out of the normalized O VI profile to produce our blend-free result. This division did not change the overall shape of the O VI line or its line center, but slightly reduced its equivalent width and hence our estimate of the O VI column density along this sight line. This deblending process reduces the observed equivalent width of the O VI λ 1031.926 line from 181 to 173 mÅ.

## 3. THE SIGHT LINE TOWARDS HD 116852

A discussion of the HD 116852 sight line is given by SS94 – we repeat some of the key points here because of their relevance to the high ion absorption in this direction.

The HD 116852 line of sight passes underneath the Sagittarius-Carina and the Norma-Centaurus spiral arms on its path through the Galactic disk and into the low halo. Courtès et al. (1970) and Bok (1971) found that the Sagittarius-Carina arm is at a distance of 1.5 kpc in the direction $l = 330°$, and studies have shown that H II regions in this arm have typical LSR velocities between –20 and –30 km s$^{-1}$ (Courtès 1972; Rickard 1974). The HD 116852 sight line passes underneath this arm at a $z$-distance of about –0.5 kpc. The second spiral arm of interest, the Norma-Centaurus arm, is located further away at a distance of 3.5 kpc in direction $l = 330°$, and is passed by the HD 116852 sight line at $z \approx -1$ kpc. The gas in this arm has velocities between –30 and –50 km s$^{-1}$ (Courtès 1972; Rickard 1974). SS94 detect a weak enhancement in Si IV and C IV at –35 km s$^{-1}$, which they suggest may be due to absorption in gas associated with the Norma arm, at $z \approx -1$ kpc. In §5.1 we discuss whether our results are consistent with this interpretation.

Different theoretical models of highly ionized gas flows in the Galactic halo predict the presence of high ions at $z$-distances of 1 kpc and above. These models include gas flows caused by Galactic fountains (Bregman 1980), Galactic chimneys (Norman & Ikeuchi 1989), and possibly shocks created by Galactic density wave perturbations (Martos & Cox 1998). Studies of high-latitude sight lines have resulted in the detection of Si IV, C IV, N V, and O VI that may have originated in Galactic supershells (Sembach & Savage 1992; Widmann et al. 1998; Savage, Sembach, & Lu 1997; Savage et al. 2000). Consequently, absorption features detected in the velocity range –20 to –30 km s$^{-1}$ or –30 to –50 km s$^{-1}$ may be associated with gas flows from spiral arms, despite the large distances from the Galactic plane. Ground-based H I 21 cm radio emission data is not readily available for HD 116852. However, Diplas & Savage (1994) analyzed the damping wings of the Lyα line using an IUE spectrum of HD 116852 to find that log N(H I) = 20.96 ± 0.08, corresponding to an average H I number density of along the line of sight of 0.06 atoms cm$^{-3}$. Evidently, at least parts of the sight line pass through low density components of the ISM. SS94 note that most of this neutral hydrogen is probably associated with gas in or near the Galactic plane, at velocities between –45 and +30 km s$^{-1}$.

Sembach, Danks, & Savage (1993) included HD 116852 in their survey of interstellar Na I and Ca II absorption and found evidence for at least five components in Na I and six in Ca II along this sight line. These authors noted that the Ca II absorption remains strong out to $v_{LSR} \approx -40$ km s$^{-1}$, which implies $z \approx -1$ kpc if the halo gas co-rotates with gas in the disk. In contrast, the Na I absorption is weaker at such velocities and distances, indicating a difference in the scale heights between the two species. HD 116852 was also one of 12 stars whose ultraviolet spectra were analyzed by Sembach & Savage (1992) using high-dispersion *IUE* observations. The low ionization features of diffuse clouds along this sight line are discussed by Sembach & Savage (1996).

In addition to passing underneath the Sagittarius–Carina arm and the Norma-Centaurus arm, the HD 116852 sight line also passes through the southern part of Radio Loop I. Thought to be a volume of hot gas created by supernovae (Iwan 1980) and strong stellar winds, Radio Loop I is a region of enhanced radio continuum emission and soft X-ray emission centered around the Sco-Cen star cluster. Analysis of early X-ray data (Nousek et al. 1982; McCammon et al. 1983) suggested that HD 116852 is located on the edge of an enhancement seen in the M band (range 0.4 to 1.1 keV), and SS94 speculated that some of the high ion absorption along this sight line could be associated with the M-band emitting gas. However, the higher angular resolution ROSAT data published by Snowden et al. (1995) reveal that the HD 116852 direction contains a low X-ray intensity in both the 1/4 and 3/4 keV bands. Studies of the absorption line spectrum of the extragalactic source 3C 273 have revealed that sight lines passing through Radio Loop I have larger C IV column densities than those present in other sight lines (Sembach, Savage, & Tripp 1997; Burks et al. 1991; Savage et al. 1993).

Although 14 H II regions lying at negative latitudes between $l = 300°$ and $l = 310°$ have been cataloged (Marsálková 1974), SS94 argue that such regions do not contribute to the high ion absorption along the HD 116852 sight line, since the ionized regions would have to be much larger than the observed radio and optical regions to intercept the sight line.

Finally, the work of Koo et al. (1992) cataloging Galactic "worms" indicates that there are several features near the Galactic plane in the direction toward HD 116852. Galactic worms are likely produced when an expanding superbubble breaks through the thin gaseous disk and forges holes in the medium extending into the halo. Such structures are observed in both H I 21 cm radio emission and in infrared emission from dust at 60 and 100 μm. SS94 report that the closest cataloged superbubble known to the HD 116852 sight line is GW 304.6 –7.8, at an approximate distance of 10 kpc, which places it far beyond HD 116852. Also, it is confined (in both IRAS 100 μm and H I 21 cm emission) to latitudes $b > -10°$, whereas HD 116852 lies at $b = -16.1°$.

## 4. RESULTS AND ANALYSIS

### 4.1 *Line Profiles and Apparent Column Densities*

Figure 1 shows plots of observed intensity against LSR velocity for the following lines: Si IV λλ1393.755, 1402.770, C IV λλ1548.195, 1550.770, N V λλ1238.821, 1242.804 and O VI λ1031.926. The four species have been displayed in this order to reflect the increasing ionization potential of each ion from the top to the bottom of the figure. The component structure within the Si IV and C IV lines is immediately apparent, whereas the N V line profiles (observed at the same resolution) contain fewer features. Although the O VI line is observed at a lower resolution, we still expect its profile to be intrinsically smoother than the C IV and Si IV line profiles (see §4.3). The ease with which continua could be fitted varied with each line: the C IV, Si IV, and O VI profiles were fairly simple to treat, whereas with N V the situation was more complicated because of the presence of a broad stellar photospheric component on which the interstellar absorption is superimposed. This same problem is discussed by SS94 with regard to the N V line in the GHRS spectrum of HD 116852. To deal with this, a continuum was estimated for the two N V lines by assuming they extend over the same LSR velocity range as the C IV and Si IV lines (–80 to 55 km s$^{-1}$). The equivalent widths measured here for the N V doublet are 43.5 ± 5.5 and 28.7 ± 4.5 mÅ, as opposed to 59.2 ± 9.0 and 30.7 ± 6.2 mÅ by SS94. We note that this difference may be partly attributed to continuum placement uncertainties, which are not fully accounted for in the quoted errors. The continuum fits adopted for each line, produced by fitting a low-order (n ≤ 5) Legendre polynomial to the data points on either side of the line profile, are displayed as short dashed lines on Figure 1. The spectrum showing the C IV 1550.770 Å line has a cut-off at around 100 km s$^{-1}$ because it was observed at the edge of an 8 Å echelle order on the STIS detector. Note the presence of the two H$_2$ lines at $v_{LSR} = -213.5$ km s$^{-1}$ and 124.9 km s$^{-1}$ in the O VI rest frame. In Figure 1 the observed O VI profile is displayed, without removal of the HD 1031.912 Å line.

Continuum normalized absorption profiles are shown in Figure 2. The O VI profile in this figure has been corrected for blending with HD λ1031.912, according to the procedure described in §2.2.

Table 4 contains the measured high ion absorption line equivalent widths, together with associated ± 1σ uncertainties due to continuum placement and statistical errors. Our statistical errors were calculated differently for the two data sets. For the STIS line profiles, we measured the r.m.s. scatter in the data around the fitted continuum on either side of the line, and then interpolated to estimate the error on each data point within the line profile, assuming Poisson statistics. This produced an error array that our equivalent width measuring software used to produce a statistical error estimate on $W_\lambda$, which is then added in quadrature with a continuum placement uncertainty to produce our quoted errors. For the FUSE line profile, the CALFUSE pipeline produces a statistical error array; again, this was added in quadrature with a continuum placement uncertainty to give our total error estimate (see Sembach & Savage 1992 for a description of this technique). The final column in Table 4 contains a measure of the signal-to-noise ratio in the continuum adjacent to each line. In order to derive a physically meaningful quantity from our empirical line profiles, we then calculate apparent column densities using the following formulae:

$$\tau_a(v) = \ln [1/I(v)] \tag{1}$$

$$N_a(v) = (m_e c/\pi e^2)[\tau_a(v)/f\lambda] = 3.768 \times 10^{14} [\tau_a(v)/f\lambda] \tag{2}$$

Equation (1) expresses $\tau_a(v)$, the apparent optical depth of the line, as a function of velocity, where $I(v)$ is the continuum normalized line profile. Equation (2) gives $N_a(v)$, the apparent column density per unit velocity, in units of ions cm$^{-2}$ (km s$^{-1}$)$^{-1}$, where $\lambda$ is the wavelength measured in Å and $f$ is the oscillator strength of the transition. This is an "apparent" column density because the line has been blurred to some extent by the instrument. Equations (1) and (2) were used to produce apparent column density profiles for the high ions, as shown in Figure 3, where for each ion we show the results from both members of the doublet. The open circles represent the stronger of the doublet lines, and the filled circles represent the weaker. The error bars in Figure 3 were produced by propagating the statistical and continuum placement errors on the flux profiles through to column density space. By integrating the apparent column density over the full extent of the line, one can determine the total column density of a given species; this method was used to produce the log $N_a$ values in Table 5.

As discussed by Savage & Sembach (1991), the integrated apparent column density, $N_a$, of a line is equivalent to the true integrated column density, $N$, if no unresolved saturated structure is present in the line profile. Such saturation can be identified by comparison of lines from a given ionic species with different $f$-values; if saturation is present, the weaker lines will show higher integrated $N_a$ values than the stronger lines. No evidence for such saturation is seen when comparing the weak and strong lines of the C IV and Si IV doublets. Our adopted $N$ values are therefore the average of the $N_a$ determinations of the weak and strong lines for these species. The continuum placement uncertainties for the N V doublets make such a comparison untenable. We therefore adopt the integrated column density of the stronger transition as our best estimate for $N$(N V).

### 4.2 *Profile Comparisons*

In Figure 4 we compare plots of normalized apparent column density as a function of velocity for the different high ions. The profiles were normalized to their peak values so that we could compare the shapes from different species more easily. In each case we display the profiles of the stronger member of the doublet. The top panel shows the profiles of C IV and Si IV, which follow each other fairly closely, with components seen at similar velocities. The positive velocity shoulder is stronger in C IV than the corresponding feature in Si IV.

In the center panel of Figure 4 we compare the C IV profile to that of O VI. The most significant difference between the profiles of C IV and O VI is the overall displacement of the O VI profile to more positive velocities than the C IV profile. Both profiles extend to roughly the same negative velocity ($\approx$ –70 km s$^{-1}$), but whereas the O VI profile can be modeled by a single Gaussian centered at –2.3 km s$^{-1}$, the C IV has a broad component centered at –21.1 km s$^{-1}$ (see §4.4). Despite the difficulties involved in obtaining an accurate velocity calibration for the O VI line, we are confident that the displacement in velocity between O VI and C IV is real.

The difference between the O VI and N V profiles is less pronounced, as seen in the lower panel of Figure 4, although the large uncertainties in the N V data make direct comparisons more difficult. Gaussian component fitting (§4.4) reveals the center of the N V absorption to be at –15.6 km s$^{-1}$, over 13 km s$^{-1}$ more negative than the broad O VI absorption. The widths of these two components are similar, with $b$(N V) = 34.1 ± 2.1 km s$^{-1}$ and $b$(O VI) = 39.4 ± 0.2 km s$^{-1}$. We discuss possible explanations for the differences between the observed profiles in §5.

### 4.3 *Relationship between High Ion, Intermediate Ion, and Low Ion Absorption*

To illustrate the difference between the profiles of cool, warm and hot gas we compare the apparent column density profiles of a low ion (Mg II), an intermediate ion (Al III, using intermediate resolution GHRS data), and a high ion (Si IV) along the HD 116852 sight line in Figure 5. In each panel data from both members of the resonance doublet are displayed. The Mg II and Si IV observations have a resolution of 2.7 km s$^{-1}$ compared to 11 km s$^{-1}$ for the Al III observations. We note the marked differences between the three panels, in terms of line shape, line width, and central velocity of absorbing components. The low ion (Mg II) has a strong component close to 3 km s$^{-1}$ (as did the Cl I and H$_2$ lines used for velocity calibration), arising from nearby gas in the Galactic disk. The intermediate ion (Al III) displays a strong component centered near –15 km s$^{-1}$, but also shows a shoulder near –40 km s$^{-1}$, and a weak extension of absorption to +30 km s$^{-1}$. It is likely that both the –40 km s$^{-1}$ and the –15 km s$^{-1}$ absorbers are associated with the strong features seen at the same velocities in Si IV (bottom panel).

In their study of gas and dust abundances in diffuse halo clouds in the Milky Way, Sembach & Savage (1996, hereafter SS96) obtained high-resolution (FWHM = 3.5 km s$^{-1}$) spectra of the low ions Fe II, Mn II, and Ca II along the HD 116852 sight line. Eight absorption components, centered at LSR velocities of –55, –42, –17, –3, 3, 8, and 18 km s$^{-1}$ were found in these three low ions using Gaussian component fitting software similar to that employed in our investigation. Interestingly, no components were found at either –36 or –10 km s$^{-1}$, where we see narrow Si IV and C IV absorption components. Indeed, inspection of Figure 3 in SS96 shows that the low ion absorption is weakest at the two places where the narrow Si IV and C IV absorption is strongest. It is possible that the broad C IV and Si IV absorption centered near at 17 km s$^{-1}$ is associated with the very weak low ion component at 18 km s$^{-1}$. That low ion component contains very little of the low ionization gas along the line of sight. In the case of Mn II the column density in the 18 km s$^{-1}$ component is 29 times smaller than in the strongest Mn II component at 3 km s$^{-1}$ (SS96). There is no evidence for broad absorption in the low ions. Overall, it is clear that the high ion structure along this sight line is markedly different from the low ion structure. The low ions are known to be confined to the Galactic plane, so the strong low ion components seen in the SS96 data set likely arise close to us in the disk of the Galaxy, whereas the high ion absorption likely traces gas extending out into the halo.

### 4.4 *Component Fitting*

We used Gaussian optical depth fitting software to simultaneously fit the high ion normalized flux profiles with a superposition of individual components. The simultaneous fitting process calculates those individual components that, after convolution with a specified instrumental spread function, combine to produce the resultant model that best fits the data. This procedure works by seeking to minimize the residuals within a three-dimensional parameter space; these three parameters are the component central velocity ($v_{LSR}$), line width (*b*), and column density (*N*). When fitting the STIS profiles we used FWHM = 2.7 km s$^{-1}$, and for the FUSE profile we used FWHM = 20 km s$^{-1}$, where FWHM is the full-width at half-maximum intensity of the Gaussian instrumental line spread function.

This simultaneous fitting procedure was applied to both the Si IV and the C IV doublets, but we were not able to satisfactorily fit the N V doublet. Consequently, when considering N V, we use the results obtained from fitting the stronger line ($\lambda$1238.821), which is better determined than the weaker line ($\lambda$1242.804). An independent fit to the stronger line was also the technique used to model the O VI doublet, since the weaker line was severely blended. In Figure 6, these component fits are displayed graphically; the circles represent data points, the light curves represent the individual line components, and the heavy curves represent the combined resulting models. In our fitting process we chose the minimum number of components that give an acceptable fit. These fits are not unique, since the observed profiles can be reproduced by a larger number of possible individual components.

The best fit parameters are given in Table 6, where for each line we include an upper limit for the temperature of the absorbing gas, calculated using the formula $T_{max}(K) = A[7.75b(\text{km s}^{-1})]^2$, where *A* is the atomic weight of the ion and *b* is the Doppler parameter obtained from the fit. To obtain a more rigorous upper limit, we used the ($b + 1\sigma$) value when computing $T_{max}$. This is an upper limit because it assumes pure thermal Doppler broadening – in reality the gas is likely to be at a lower temperature, with non-thermal (kinematic) mechanisms providing a significant contribution to the line broadening. By comparing the observed widths of the components with temperature estimates derived from ionic ratios, one can constrain the magnitude of the non-thermal contribution to the line broadening.

The profiles indicate that three types of high ionization absorbing gas are present along this sight line. First, we find a broad absorption component, spanning a range of positive and negative velocities, and present in the profiles of all four ions analyzed here; we have labeled this absorption feature Component 1. Second, we detect two narrow absorption components in the Si IV and C IV profiles, labeled Components 2 and 3; such absorption is not apparent in the N V and O VI data. Finally, the Gaussian fitting procedure identified an intermediate-width absorption component in the profiles of Si IV and C IV (Component 4). This component could be classified as narrow, but its width may be suggestive of a different ionization mechanism than that occurring in Components 1 and 2. The origin and ionization of these different types of gas forms a large part of the discussion in §5.

### 4.5 *Component-to-Component Line Ratios*

For each of the components labeled 1 to 4 in Table 6 we can calculate ratios between the absorbing species. Such column density ratios are useful because not only can they be compared between sight lines, but they can also be compared with predictions from theoretical ionization models. For the ions Si IV and C IV we compute the column density ratios within the two narrow components (Components 2 and 3) and additionally within the intermediate-width component (Component 4). Ratios for all four ions studied here are calculated within the broad absorbing component (Component 1). The results are shown in Table 7, and are discussed more in the context of ionization in §5.2. The quoted errors were calculated by propagating the 1σ error on N returned by our simultaneous component fitting software.

### 5. GAS KINEMATICS AND IONIZATION

Along the 4.8 kpc path to HD 116852, Galactic rotation affects the negative velocity extension of the absorption profiles we observe. These profiles are also affected by the thermal Doppler and turbulent velocity dispersion of individual species, the stratification of gas perpendicular to the plane, elemental depletion into dust, local ionization mechanisms, and the influence of large-scale structures, such as spiral arms. High-resolution absorption line spectroscopy provides the ideal opportunity to investigate these processes, and in this section we discuss the factors that can account for the different structure seen in the high ion profiles towards HD 116852.

*5.1 Gas Kinematics*

If the gas in the halo of the Milky Way co-rotates with the underlying disk, then a simple rotation law can be used to convert a measured line-of-sight velocity to a distance estimate for an absorbing cloud. In this way we can build up information concerning the physical location of clouds in the Galactic disk and halo. The assumption of co-rotation is justified by the work of Lockman (1984), who reported on studies of extended H I in the Galaxy between radii of 4 and 8 kpc, and found that this gas co-rotates up to about 1 kpc away from the plane. For the high ions, work by Savage, Massa, & Sembach (1990) and Sembach, Savage, & Massa (1991) has shown that the halo gas near the Galactic center can decouple from the underlying disk gas, but this occurs at $z$-distances greater than 1.3 kpc. Furthermore, the HD 116852 sight line does not pass close enough to the Galactic center where this disk/halo decoupling seems to exist. We note that Galactic fountain models of gas flow into the halo produce motions that could be interpreted as departures from co-rotation. Such fountain motions directed away from and towards the plane could result in peculiar cloud motions in these directions. However, the line-of-sight component of motion perpendicular to the Galactic plane is reduced by a factor of sin |$b$| (i.e., for the HD 116852 by 0.28), which serves to lessen the impact of such motions on the observed line profiles.

Sembach & Savage (1996) analyzed the low ionization states and chemical abundances along this sight line with GHRS data, and identified the following three absorption intervals, defined by their LSR velocities:

Group 1:    $v_{LSR} \geq -10$ km s$^{-1}$

Group 2:    $-35 \leq v_{LSR} \leq -10$ km s$^{-1}$

Group 3:    $-70 \leq v_{LSR} \leq -35$ km s$^{-1}$

If the gas in the halo is assumed to co-rotate with the Galactic disk out to the $z$-distance of HD 116852 (–1.3 kpc), then according to the Clemens (1985) rotation curve these velocities correspond to the following intervals in distance, $z$-distance, and Galactocentric distance:

Group 1:  $0.0 \leq d \leq 0.7$ kpc; $0.0 \leq |z| \leq 0.2$ kpc; $8.5 \geq r_G \geq 8.1$ kpc

Group 2:  $0.7 \leq d \leq 2.7$ kpc; $0.2 \leq |z| \leq 0.7$ kpc; $8.1 \geq r_G \geq 7.3$ kpc

Group 3:  $2.7 \leq d \leq 4.8$ kpc; $0.7 \leq |z| \leq 1.3$ kpc; $7.3 \geq r_G \geq 7.0$ kpc

Sembach & Savage (1996) suggest that absorption in Groups 2 and 3 is probably associated with halo gas flowing out of the Sagittarius-Carina and Norma-Centaurus spiral arms, respectively. On the basis of our STIS observations of Si IV and C IV, we would agree with this assessment; the features we identified as Components 2 and 3 correspond to absorption in Groups 3 and 2, respectively.

The intermediate-width component of absorption we detect in the Si IV and C IV profiles (Component 4) occurs at positive velocity (near 17 km s$^{-1}$), and so cannot be accounted for by co-rotating gas, since along this sight line co-rotation causes absorption at negative velocities only. As discussed in §4.3, we consider it plausible that this 17 km s$^{-1}$ absorber is associated with very weak low ion absorption at 18 km s$^{-1}$. If this low ion feature arises nearby, one might expect the Si IV and C IV absorption to be tracing an interface between the cool/warm gas and the ambient hot medium. However, without independent information about the velocity structure of gas along this line of sight as a function of distance, we have no way of determining the distance to this absorber.

Surveys of high ion absorption over a large sample of sight lines have recently been used to estimate the scale heights and mid-plane densities of the high ions perpendicular to the Galactic plane (Savage et al. 1997, 2000). These scale height ($H$) and mid-plane density ($n_0$) estimates are summarized in Table 8. We have computed the expected column density of each ion by integrating the exponential density distribution, i.e. $n(z) = n_0 e^{-(|z|/H)}$, over the length of the sight line. This model value, calculated using $N_{mod} = n_0 H(1 - e^{-|z|/H})/\sin|b|$ is given in the sixth column of Table 8. For comparison, the observed column densities are listed in the seventh column. The small discrepancies between the model and the observations (< 0.1 dex) confirm the validity of the exponential model and the current scale height and density estimates as a descriptor of the distribution of high ions over the small range in $z$ probed by the HD 116852 sight line.

Using the scale height arguments, we can also calculate the predicted value of $n(z)$ for any of the four high ions at any given distance from the plane. This method was used to calculate the values in the final column of Table 8, which are the predicted number densities of each of the high ions at −1.3 kpc, the $z$-distance of HD 116852. It can be seen that the variation in Si IV and C IV number densities from the mid-plane up to 1.3 kpc (i.e., the difference between columns 4 and 8) is much less than the variation in N V and O VI over 1.3 kpc. This is due to the observed trend from in Si IV to O VI in which the species with higher ionization potentials have *smaller* scale heights. Furthermore, we can use this density profile to predict what the $N$(C IV)/$N$(Si IV) ratio should be at the points where the HD 116852 sight line passes underneath the Sagittarius-Carina and Norma-Centaurus spiral arms, i.e., at –0.5 kpc and –1.0 kpc. These predictions can be compared with our measured ionic ratios in Components 2 and 3. The simple model predicts $N$(C IV)/$N$(Si IV) = 3.9 at –0.5 kpc where our observed ratio is 0.7 ± 0.4 (Component 3), and $N$(C IV)/$N$(Si IV) = 3.8 at –1.0 kpc where we measure 1.4 ± 0.3 (Component 2). The differences between our observed values and the simple model prediction highlight the inhomogeneity of the gas distribution in the halo; in other words, although the simple exponential distribution is followed on a large scale, there exist local regions where particular species may be over- or under-abundant. This sight line evidently samples regions where the $N$(C IV)/$N$(Si IV) ratio is unusually low. The ionic ratios are discussed further in §5.2 with regard to ionization processes.

One of the most intriguing results of our investigation is the difference between the central velocities of the C IV and O VI broad absorption features. Whereas the broad C IV absorption is centered at –21.1 km s$^{-1}$, the broad O VI absorption is centered at –2.3 km s$^{-1}$. This effect may be partially explained by the different scale heights of C IV and O VI in a co-rotating Galactic halo. As the scale height of an ionic species increases, the observed absorption line profile along the HD 116852 sight line shifts to become centered at more negative velocities. We refer the reader to Figure 7 of SS94 for a graphical illustration of this point. In this case, the scale height of C IV is 4.4 kpc and the scale height of O VI is 2.7 kpc, but we doubt this can explain the observed shift of 19 km s$^{-1}$ in the central velocity of absorption. Therefore, we suggest that either two kinematically distinct clouds are responsible for the broad C IV and O VI absorption, or that some physical process (e.g., a shock wave) is displacing in velocity the absorption by different species at the same location. However, the fact that N V is only displaced by 6 km s$^{-1}$ from the broad C IV absorption, whereas the O VI is displaced by 19 km s$^{-1}$ makes the second interpretation hard to justify. In other words, it seems difficult to visualize a mechanism which would serve to displace in velocity the O VI absorption from the C IV absorption within one location, without also affecting N V in a similar manner. Therefore, we advance the possibility that the broad C IV and O VI absorption could be occuring in *physically different locations*. If this is indeed the case, then clearly the gas containing the O VI will be at a higher temperature than the gas containing the C IV (and the other high ions). The collisional ionization equilibrium calculations of Sutherland & Dopita (1993) predict that at temperatures in the range $5.5 < \log T < 6.0$, the number density of O VI should be around two orders of magnitude greater than the number density of C IV or N V, assuming solar abundances. If a region of gas exists within this temperature range, then we would detect O VI, but not C IV or N V, just as observed at $-2.3$ km s$^{-1}$. Conversely, at temperatures of $\log T < 5.2$, the abundance of O VI drops off sharply whereas the C IV abundance peaks. Therefore, gas existing in this temperature range at a velocity of $-21.1$ km s$^{-1}$ would account for the lack of detection of O VI at the velocity where C IV is centered. If the C IV and O VI are tracing different interstellar regions, then they are tracing regions of different temperature. We note that in reality, rapid cooling may render the collisional ionization equilibrium assumption invalid, but it is clear that O VI can reside in higher temperature regions than C IV.

### *5.2. Gas Ionization*

#### 5.2.1 Narrow Components of Absorption

For both lines in the Si IV and C IV doublets, two narrow components are seen, centered at approximately $-36$ and $-10$ km s$^{-1}$. These components are more pronounced in Si IV than in C IV, and are not present at all in N V and O VI, for which we observe only one broad component. Measurements of the widths of the narrow components in Si IV and C IV suggest that photoionization is the likely ionization mechanism, because the derived upper-limit temperature estimates in the narrow components (2 and 3) of Si IV and C IV are of the order $(2 - 10) \times 10^4$ K, with the stronger constraint $(2 - 3) \times 10^4$ K provided by C IV. Without an ionizing radiation field, little Si IV and C IV would be present at these temperatures if collisional ionization equilibrium exists (Sutherland & Dopita 1993), indicating that there must be a flux of photons maintaining the observed high ionization states. However, if non-equilibrium conditions exist, then it is possible that Components 2 and 3 are collisionally ionized, given that interstellar gas at temperatures near $10^5$ K cools more rapidly than it recombines, so that "frozen-in" ionization can remain at temperatures well below those at which the equilibrium abundances peak (Kafatos 1973; Shapiro & Moore 1976). For example, non-equilibrium time-dependent shock-heated cooling models predict that the ions Si IV and C IV can exist at temperatures as low as $10^4$ K (Shapiro & Moore 1976; Edgar & Chevalier 1986). Although this non-equilibrium collisional ionization theory cannot categorically be ruled out, we favor photoionization as the likely ionization mechanism for Components 2 and 3.

Furthermore, we suggest that hot stars are the primary source of the ionizing radiation, because of photon energy arguments. The outward flux of photons from hot stars diminishes sharply above 54 eV, due to the He II ionization edge; therefore, Si IV (33.5 eV) and C IV (47.9 eV) can be produced by hot star photons, whereas N V (77.5 eV) and O VI (113.9 eV) cannot. Extragalactic radiation from AGNs does not have such a sharp He II edge, and so would be expected to photoionize N V and O VI; these species are not seen in Components 2 and 3. Photoionization models in the Galactic halo have found that the dominant source of ionizing photons depends on the photon energy. For example, Bregman & Harrington (1980) found that OB star associations in the disk dominate the ionizing radiation field in the range 13.6 to 45 eV, planetary nebulae provide the majority of the photons between 45 and 54 eV, and AGN and the Galactic soft X-ray background provide most of the photons above 54 eV. These results suggest that the source of ionizing photons in the narrow components is mainly O and B stars, with perhaps some contribution from planetary nebulae.

Of particular interest is the $N$(C IV)/$N$(Si IV) ratio in the two narrow components, where its value is 1.4 ± 0.3 and 0.7 ± 0.4 respectively. These values are low by interstellar standards, particularly in Component 3, thought to arise underneath the Sagittarius-Carina spiral arm. Sembach & Savage (1992) found an average Galactic value <$N$(C IV)/$N$(Si IV)> = 3.6 ± 1.3, confirming the contention of Pettini & West (1982) that this ratio is fairly constant over most Galactic sight lines. However, this Galactic average was derived from medium-resolution surveys that were unable to resolve narrow components. Consequently, it remains possible that the observed ionic ratios in Components 2 and 3 are not unusually low for narrow components. SS94 also reached the conclusion that the $N$(C IV)/$N$(Si IV) ratio is low along this sight line, although again their estimate of the ratio (2.8 to 4.5) is obtained from data which does not resolve the component structure.

If Components 2 and 3 in Si IV are produced in the same gas as Components 2 and 3 in C IV, we can constrain the non-thermal broadening present in these components by insisting that the Si IV and C IV are at the same temperature. For Si IV, $T_{max}$ in Components 2 and 3 is $1.0 \times 10^5$ and $0.57 \times 10^5$ K, and for C

IV, the corresponding values are $0.34 \times 10^5$ and $0.20 \times 10^5$ K. If the non-thermal broadening component $b_{nt}$ and the temperature were the same for both Si IV and C IV, then the Si IV component would be *narrower* than the C IV component, due to the difference in atomic weight ($A$(Si IV) = 28, $A$(C IV) = 12). However, the opposite trend is observed in our data, since we find $b$(Si IV) = 7.4 ± 0.5 km s$^{-1}$ and $b$(C IV) = 5.9 ± 1.0 km s$^{-1}$ in Component 2, and $b$(Si IV) = 5.2 ± 0.6 and $b$(C IV) = 3.6 ± 1.7 in Component 3 (although the values agree within the $2\sigma$ error range). Although the errors on these widths are fairly large, inspection of Figures 1 and 2 shows that the narrow components of Si IV really do appear to be wider than those in C IV, and so the more likely explanation (provided $T$ is the same) is that $b_{nt}$ must be different for the two species.

The magnitude of the non-thermal broadening $b_{nt}$ is related to the observed width $b$ and gas temperature $T$ according to $b_{nt} = (b^2 - 2kT/m)^{1/2}$, where $m$ is the mass of the ion, and $k$ is the Boltzmann Constant. If the gas in Component 2 were at $10^4$ K, the thermal broadening $(2kT/m)^{1/2}$ is 2.4 km s$^{-1}$ for Si IV and 3.7 km s$^{-1}$ for C IV, and therefore the required non-thermal broadening is $b_{nt}$ = 7.0 km s$^{-1}$ for Si IV, and 4.6 km s$^{-1}$ for C IV. A greater turbulent broadening for Si IV compared to C IV in photoionized gas seems plausible since Si IV is easier to ionize than C IV. So, if the harder photons necessary to photoionize C IV were localized to small regions, whereas the less energetic photons still able to photoionize Si IV occupied a more extended volume, then the observed kinematic broadening would be higher for Si IV.

### 5.2.2 Intermediate-Width Components of Absorption

Component 4, observed in the profiles of Si IV and C IV at 17 km s$^{-1}$, has a velocity very similar to the very weak low ionization component (labeled 1a) seen by SS96 in Fe II, Mn II, and Ca II. That low ionization absorption component has $N$(Mn II) = (6.5 ± 2.6) × $10^{11}$ cm$^{-2}$ and $b$(Mn II) = 5.1 ± 1.7 km s$^{-1}$. If the Mn has normal depletion properties for low column density warm neutral clouds in the disk (Sembach & Savage 1996), the amount of warm neutral hydrogen associated with the cloud is small ($\approx 1.6 \times 10^{19}$ cm$^{-2}$). Given that the high ionization absorption is directly associated in velocity with this low ionization cloud, but is much broader ($b$(Si IV) = 12.9 ± 2.0 km s$^{-1}$ and $b$(C IV) = 10.6 ± 3.3 km s$^{-1}$), it is plausible that the high ionization species arise in the conductive interface between the warm cloud and a much hotter surrounding medium.

Intermediate-width absorption line features similar to Component 4 have been observed in previous high-resolution studies of the ISM in the Galactic disk and halo. For example, along the ζ Oph sight line, Sembach, Savage, & Jenkins (1994) find Si IV and C IV absorption components with $b = 15 - 18$ km s$^{-1}$. They propose this absorption arises in a conductive interface, either on the front and rear surface of a cool cloud embedded in the local ISM, or on the edge of the expanding stellar wind-driven bubble surrounding ζ Oph. In the former case, the component would arise at a distance of less than 60 pc. Additionally, in the case of the HD 167756 sight line, Savage, Sembach, & Cardelli (1994) conclude that conductive interfaces provide the most likely explanation for structured Si IV and C IV absorption seen at positive velocities in this direction. They favor this interpretation over photoionization, since in the latter case a remarkably uniform radiation field is required over large distances to explain the observed constancy of the $N$(C IV)/$N$(Si IV) ratio.

### 5.2.3 Broad Components of Absorption

The broad absorption components, seen in all four species analyzed here, have widths that are consistent with high temperatures and hence ionization by collisions with electrons. These broad components display a trend in which the central velocity moves to positive velocity through the sequence Si IV, C IV, N V, to O VI. Scale height arguments seem unlikely to explain such a large shift in central velocity (§5.1), suggesting that the O VI resides in a separate, higher temperature phase of gas than the collisionally ionized Si IV and C IV. If these ions are produced in physically different regions, then any ionic ratios calculated between C IV and O VI are misleading, since such ratios are only of value when comparing number densities *within a single cloud.*

In the broad components ratios between all four atomic species studied here can be calculated. We measure $N$(C IV)/$N$(Si IV) = 4.3 ± 0.3. This ratio is harder to predict theoretically than $N$(C IV)/$N$(N V) or $N$(C IV)/$N$(O VI), largely because photons emitted by the cooling hot gas may well be able to re-ionize Si IV, and the exact nature of this self-photoionization is not well known. Our observed value of $N$(C IV)/$N$(Si IV) is similar to estimates of the Galactic average present in the literature: Sembach & Savage (1992) measured a value of 3.6 ± 1.3 over a number of halo sight lines, and Sembach, Savage, & Tripp (1997) measured a value of 3.8 ± 1.9 over a number of disk and halo sight lines. Our value of $N$(C IV)/$N$(Si IV) = 4.3 ± 0.3 is also consistent with one-dimensional planar flow models that account for self-photoionization (Shapiro & Benjamin 1993).

The ratio of $N$(C IV)/$N$(N V) in the broad component, measured here as 4.8 ± 0.4, is statistically identical to the Galactic average value of 4.6 ± 2.3 (Sembach & Savage 1992). Our value also agrees with that reported for this sight line by SS94, who found 4.3 ± 0.8, even though they did not resolve the component structure, and used a different continuum placement for the N V line. If the gas producing these broad components is in collisional ionization equilibrium, then the observed ionic ratio implies a kinetic temperature of $\approx 1 \times 10^5$ K (Sutherland & Dopita 1993). Furthermore, if this assumption of collisional ionization equilibrium is valid, then we can infer that there must be a significant turbulent broadening component in the line profiles, since in the case of pure thermal broadening the observed C IV and N V line widths correspond to temperatures of $9 \times 10^5$ and $10 \times 10^5$ K respectively.

We now turn our attention to the $N$(C IV)/$N$(O VI) ratio in the broad component, which we measure to be 0.54 ± 0.02. This ratio only has physical significance if the C IV and O VI co-exist in the same absorbing medium. For the sake of argument, we assume that the two species do reside together, so as to examine what processes could produce the measured ionic ratio. The observed value for the ratio is significantly lower from that predicted by the turbulent mixing layer calculations performed by Slavin, Shull & Begelman (1993), whose models predict a value of 1.4 if the mixing layer has log $T_m$ = 5.0, and a value of 8 if log $T_m$ = 5.5, where $T_m$ is the temperature of the mixing layer. The observed ratio is easier to reconcile with a radiative cooling model, in which different ionization stages are passed through as a volume of gas cools down. Shapiro & Benjamin (1993) have used such a model applied to the cases of isochoric and isobaric gas flow, predicting $N$(C IV)/$N$(O VI) = 0.3 and 0.5 respectively. We note that our observed ratio is also similar to that predicted by various conductive heating models, in which heat flows across the conduction front between hot gas and a cooler cloud. For example, the model of Borkowski, Balbus, & Fristrom (1990) considers a plane interstellar cloud threaded by a strong magnetic field and embedded within a hot, uniform medium. Their calculations predict $N$(C IV)/$N$(O VI) ratios that depend on both time and the angle $\theta$ between the magnetic field and the $z$-axis. For a value $\theta = 0°$, the equilibrium ratio is 0.25, but this falls as the magnetic field component in the plane of the disk increases. Slavin & Cox (1992, 1993) investigated the conductive heating associated with the evolution of supernova remnants, in which a cold postshock shell evaporates into the expanding bubble. Their prediction for $N$(C IV)/$N$(O VI) is between 0.12 and 0.42, depending on a number of model parameters. We refer the reader to Spitzer (1996) for a succinct summary of the $N$(C IV)/$N$(O VI) ratio predictions for a variety of radiative cooling, conductive interface, and turbulent mixing layer models present in the literature.

Shull & Slavin (1994) made the suggestion that whereas halo high ion abundances are more easily explained by turbulent mixing layer (and superbubble) processes, disk high ion abundances are more consistent with conductive interface and cooling supernova remnant models. This hybrid picture, in which the halo ionization mechanism is different from the disk ionization mechanism, was also suggested by Sembach (1994). With regard to the data set analyzed here, the fact that TML models significantly overestimate the observed $N$(C IV)/$N$(O VI) ratio suggests that maybe the broad O VI absorption does not occur in the halo, but rather nearer to us in the disk. This might also explain the observed displacement of ~20 km s$^{-1}$ between the line centers of the ions C IV and O VI. Although the Local Bubble is known to be at temperatures hot enough to produce O VI, no evidence exists for significant column densities of O VI to reside within the Local Bubble itself (Oegerle et al. 2000). Therefore, although we cannot localize the broad O VI component, it is almost certainly produced no closer than the edge of the Local Bubble.

## 6. FREQUENCY OF OCCURENCE OF HIGH IONIZATION ABSORPTION COMPONENTS

The frequency of absorbing components along this sight line is comparable to that seen in other sight lines studied at high resolution. Along the 4.8 kpc path to HD 116852, we detect four C IV components (two narrow, one intermediate-width and one broad), corresponding to a frequency of 0.83 kpc$^{-1}$. Savage, Sembach, & Howk (2001) found four C IV components over 4.9 kpc towards HD 177989 (0.82 kpc$^{-1}$); Savage, Sembach, & Cardelli (1994) detected five C IV components over 4.0 kpc towards HD 167756 (1.25 kpc$^{-1}$); Sembach, Savage, & Tripp (1997) found three C IV components over 4.1 kpc towards HD 119608 (0.73 kpc$^{-1}$); Fitzpatrick & Spitzer (1997) found four C IV components over 2.7 kpc towards HD 215733 (1.48 kpc$^{-1}$). Combining these five lines of sight we find a total of 20 C IV components over 20.5 kpc, or a frequency of 1.0 ± 0.25 kpc$^{-1}$. We note that this may be regarded as a lower limit since the fitting process is conservative in identifying distinct components.

## 7. SUMMARY

We summarize our study of high-resolution (FWHM ≈ 2.7 km s$^{-1}$) STIS E140H observations together with medium-resolution (FWHM ≈ 20 km s$^{-1}$) FUSE LiF1A observations of HD 116852 (O9 III, $l$ = 304.9°, $b$ = –16.1°, $d$ = 4.8 kpc, $z$ = –1.3 kpc) in the following key points:

1. We convert the observed absorption line profiles into profiles of apparent column density per unit velocity. By integrating over the full extent of each profile, we find that log $N$(Si IV) = 13.60 ± 0.02, log $N$(C IV) = 14.08 ± 0.03, log $N$(N V) = 13.34 $\pm_{0.06}^{0.05}$, and log $N$(O VI) = 14.28 ± 0.01.

2. By using a simultaneous component fitting routine we find a four-component structure to the C IV and Si IV profiles, with two narrow components (Components 2 and 3) centered at –36 and –10 km s$^{-1}$, one intermediate-width component (Component 4) centered at 17 km s$^{-1}$, and a broad component (Component 1) centered near –21 km s$^{-1}$. In contrast, we find only single component structures to N V and O VI, with broad features centered at –16 km s$^{-1}$ and –2 km s$^{-1}$ respectively.

3. We interpret the two narrow components (Components 2 and 3) present in Si IV and C IV as being produced in gas that is photoionized. This is because the observed line widths imply the gas is warm but not hot (i.e. $T$ ≈ few × 10$^4$ K), and (equilibrium) collisional ionization will not produce the ions Si IV and C IV at these temperatures. It is possible that non-equilibrium collisional ionization could also produce the low line widths. Furthermore, we suggest that Components 2 and 3 originate in gas outflowing from the Norma-Centaurus and Sagittarius-Carina spiral arms respectively.

4. The component of intermediate width, detected in Si IV and C IV at 17 km s$^{-1}$, may be produced at the conductive interface between a cooler cloud and the surrounding hot ISM. However, we have no information about the location of this feature.

5. The broad component present in all four species is likely produced in collisionally ionized hot gas. The $N$(C IV)/$N$(Si IV) and $N$(C IV)/$N$(N V) ratios within Component 1 are comparable to the Galactic average. The $N$(C IV)/$N$(O VI) ratio of 0.54 ± 0.02 is more consistent with radiative cooling and conductive heating than turbulent mixing layer models, although the velocity displacement of the line center between Component 1 in C IV and O VI may suggest that these ions reside in altogether different locations.

6. The frequency of occurrence of absorption components in C IV and Si IV along the HD 116852 sight line is comparable to that measured along other sight lines at comparable resolution. Data collected over five sight lines yields a mean frequency of occurrence for C IV components of $1.0 \pm 0.25$ kpc$^{-1}$.

The O VI results were obtained for the Guaranteed Time Team by the NASA-CNES FUSE Mission operated by Johns Hopkins University. Financial support to US participants has been provided by NASA contract NAS5-32985. The results for the other highly ionized species are based on observations obtained with the Space Telescope Imaging Spectrograph on the NASA/ESA Hubble Space Telescope, obtained at the Space Telescope Science Institute, which is operated by the Association of Universities for Research in Astronomy, Inc., under NASA contract NAS5-25655. KRS acknowledges financial support from NASA Long Term Space Astrophysics grant NAG5-3485.

REFERENCES

Allison, A. C., & Dalgarno, A. 1970, Atomic Data, 1, 289

Bok, B. J. 1971, in *Highlights in Astronomy*, Vol. 2, ed. C. deJager (Washington: GPO), 63

Borkowski, K. J., Balbus, S. A., & Fristrom, C. C. 1990, ApJ, 355, 501

Bregman, J. N. 1980, ApJ, 236, 577

Bregman, J. N., & Harrington, J. P. 1986, ApJ, 309, 833

Burks, G. S., York, D. G., Blades, J. C., Bohlin, R.C., & Wamsteker, W. 1991, ApJ, 381, 55

Clemens, D. P. 1985, ApJ, 295, 422

Courtès, G. 1972, Vistas Astron., 14, 81

Courtès, G., Georgelin, Y. P., Georgelin, Y. M., & Monet, G. 1970, in IAU Symp. 38, *The Spiral Structure of Our Galaxy*, ed. W. Becker & G. Contopoulos (Dordrecht: Reidel), 209

Dabrowski, I., & Herzberg, G. 1976, Canadian Journal of Physics, 54, 525

Diplas, A., & Savage, B. D. 1994, ApJS, 93, 211

Edgar, R. J., & Chevalier, R. A. 1986, ApJ, 310, L27

Fitzpatrick, E. L., & Spitzer, L. 1997, ApJ, 475, 623

Howk, J. C., & Sembach, K. R. 2000, AJ, 119, 2481

Iwan, D. 1980, ApJ, 239, 316

Jenkins, E. B. 1978a, ApJ, 219, 845

Jenkins, E. B. 1978b, ApJ, 220, 107

Jura, M., & York, D. G. 1978, ApJ, 219, 861

Kafatos, M. 1973, ApJ, 182, 433

Kimble, R. A., et al. 1998, ApJ, 492, L83

Koo, B. C., Heiles, C., & Reach, W. T. 1992, ApJ, 390, 108

Lauroesch, J. T., Meyer, D. M., Watson, J. K., & Blades, J. C. 1998, ApJ, 507, L89

Lockman, F. J. 1984. ApJ, 283, 90

Marsálková, P. 1974, Ap&SS, 27, 3

Martos, M. A., & Cox, D. P. 1998, ApJ, 509, 703

McCammon, D., Burrows, D. N., Sanders, W. T., & Kraushaar, W. L. 1983, ApJ, 269, 107

Mihalas, D., & Binney, J. 1981, *Galactic Astronomy*, 2$^{nd}$ edition (San Francisco: Freeman)

Moore, C. E. 1970, NSRDS-NBS 34

Moos, H. W. et al. 2000, ApJ, 538, L1

Morgan, W. W., Code, A. D., & Whitford, A. E. 1955, ApJS, 2, 41

Morton, D. C. 1991, ApJS, 77, 119

Norman, C., & Ikeuchi, S. 1989, ApJ, 345, 372

Nousek, J. A., Fried, P. M., Sanders, W.T., & Kraushaar, W. L. 1982, ApJ, 258, 83

Oegerle, W. R., Jenkins, E. B., Shelton, R. L., Bowen, D. V., & FUSE Team 2000, AAS 197, #07.18

Pettini, M., & West, K. A. 1982, ApJ, 260, 561

Rickard, J. J. 1974, A&A, 31, 47

Sahnow, D. J. et al. 2000, ApJ, 538, L7

Savage, B. D., Lu, L., Weymann, R., Morris, S., & Gilliland, R. 1993, ApJ, 404, 134

Savage, B. D., Massa, D., & Sembach, K. R. 1990, ApJ, 355, 114

Savage, B. D., & Sembach, K. R. 1991, ApJ, 379, 345

Savage, B. D., Sembach, K. R., & Cardelli, J. A. 1994, ApJ, 420, 183

Savage, B. D., Sembach, K. R., & Howk, J. C. 2001, ApJ, 547, 907

Savage, B. D., Sembach, K. R., & Lu, L. 1997, AJ, 113, 2158

Savage, B. D. et al. 2000, ApJ, 538, L27

Sembach, K. R. 1994, ApJ, 434, 244

Sembach, K. R., Danks, A., & Savage, B. D. 1993, A&AS, 100, 107

Sembach, K. R., & Savage, B. D. 1992, ApJS, 83, 147

Sembach, K. R., & Savage, B. D. 1994, ApJ, 431, 201

Sembach, K. R., & Savage, B. D. 1996, ApJ, 457, 211

Sembach, K. R., Savage, B. D., & Jenkins, E. B. 1994, ApJ, 421, 585

Sembach, K. R., Savage, B. D., & Massa, D. 1991, ApJ, 372, 81

Sembach, K. R., Savage, B. D., & Tripp, T. M. 1997, ApJ, 480, 216

Shapiro, P. R., & Benjamin, R. A. 1993, in *Star Formation, Galaxies and the Interstellar Medium*, ed. J. J.

Franco, F. Ferrini, & G. Tenorio-Tagle (New York: Cambridge University Press), 275

Shapiro, P. R., & Moore, R. T. 1976, ApJ, 207, 406

Shull, J. M., & Slavin, J. D. 1994, ApJ, 427, 784

Slavin, J. D., & Cox, D. P. 1992, ApJ, 392, 131

Slavin, J. D., & Cox, D. P. 1993, ApJ, 417, 187

Slavin, J. D., Shull, J. M., & Begelman, M. C. 1993, ApJ, 407, 83

Snowden, S. L., Egger, R., Finkbeiner, D. P., Freyberg, M. J., & Plucinsky, P. P. 1998, ApJ, 493, 715

Snowden, S. L., Freyberg, M. J., Plucinsky, P. P., Schmitt, J. H. M. M., Trümper, J., Voges, W., Edgar, R. J., McCammon, D., & Sanders, W. T. 1995, ApJ, 454, 643

Spitzer, L. 1996, ApJ, 458, L29

Sutherland, R. S., & Dopita, M. A. 1993, ApJS, 88, 253

Widmann, H., et al. 1998, A&A, 338, L1

Woodgate, B. E., et al. 1998, PASP, 110, 1183

TABLE 1
BASIC INFORMATION FOR HD 116852

| MK | V | E(B–V) | l | b | d | z | $r_G$[a] | log N(H I) |
|---|---|---|---|---|---|---|---|---|
| O9 III[b] | 8.47[c] | 0.22[c] | 304.9° | –16.1° | 4.8 kpc | –1.3 kpc | 7.0 kpc | 20.96[d] |

[a] $r_G$ = Galactocentric radius.
[b] Spectral Classification by Morgan, Code, & Whitford (1955), confirmed by Sembach & Savage (1994).
[c] V magnitude and color excess from Sembach et al. (1993).
[d] Neutral hydrogen column density from Diplas & Savage (1994), where N is measured in cm$^{-2}$.

TABLE 2
SPACE TELECOPE IMAGING SPECTROGRAPH OBSERVATIONS OF HD 116852

| Dataset[a] | Grating | Wavelength Coverage (Å) | Exposure Time (s) | Resolution[b] FWHM (km s$^{-1}$) |
|---|---|---|---|---|
| 05C01C010 | E140H | 1170 - 1372 | 360 | 2.7 |
| 063571010 | E140H | 1388 - 1590 | 720 | 2.7 |

FAR ULTRAVIOLET SPECTROSCOPIC EXPLORER OBSERVATIONS OF HD 116852

| Program ID | Channel/Segment[c] | Wavelength Coverage (Å) | Exposure Time (s) | Resolution[d] FWHM (km s$^{-1}$) |
|---|---|---|---|---|
| P1013801 | LiF1A | 987 - 1082 | 7212 | ≈ 20 |

[a] All STIS observations were taken with the 0.2″ x 0.2″ aperture, using the STIS/FUV-MAMA configuration in ACCUM mode.
[b] STIS velocity resolution is estimated from $R = \lambda/\Delta\lambda \approx 110\,000$ (STIS Instrument Handbook).
[c] The LiF1A channel/segment produced the highest S/N FUSE spectrum of the region near O VI λλ 1031.926, 1037.627 and was chosen as the source of data in preference to the other channels. Twelve individual extractions (six of exposure time 504 s, and six of 698 s) were co-added with no wavelength shifts in order to produce the combined LiF1A spectrum used in this paper (total on-spectrum exposure time 7212 s).
[d] FUSE velocity resolution is estimated to be between 20 and 25 km s$^{-1}$.

TABLE 3
FUSE OBSERVATIONS OF HD LYMAN SERIES LINES

| Line | $\lambda$[a] (Å) | f-value[b] | $v_{LSR}$[c] (km s$^{-1}$) | $W_\lambda$ (mÅ) |
|---|---|---|---|---|
| (8-0) R(0) | 1011.457 | 0.0244 | 3.4 | 16.1 ± 3.6 |
| (7-0) R(0) | 1021.456 | 0.0242 | 4.3 | 11.6 ± 2.7 |
| (6-0) R(0) | 1031.912 | 0.0228 | …[d] | …[d] |
| (5-0) R(0) | 1042.847 | 0.0201 | –5.0 | 18.7 ± 3.4 |
| (4-0) R(0) | 1054.288 | 0.0161 | –4.6 | 12.7 ± 3.4 |

[a] HD vacuum wavelengths taken from Dabrowski & Herzberg (1976).
[b] HD oscillator strengths based upon calculations by Allison & Dalgarno (1970).
[c] The ≈10 km s$^{-1}$ range of observed LSR velocities for these HD lines illustrates the uncertainties in the FUSE wavelength scale over the wavelength range from 1010 to 1055 Å.
[d] The measurements of the blend-free HD lines in this table are used to infer the 15.6 mÅ strength of HD (6-0) R(0) λ1031.912 that blends with O VI λ1031.926 (see §2.2).

TABLE 4
EQUIVALENT WIDTHS OF HIGH ION ABSORPTION LINES

| Ion | $\lambda$[a] (Å) | $f$-value[a] | $W_\lambda$[b] (mÅ) | Integration Range (km s$^{-1}$) | S/N[c] |
|---|---|---|---|---|---|
| Si IV | 1393.755 | $5.140 \times 10^{-1}$ | $223.3 \pm 3.6$ | −80 to +55 | 25 |
| Si IV | 1402.770 | $2.553 \times 10^{-1}$ | $131.3 \pm 4.5$ | −80 to +55 | 32 |
| C IV | 1548.195 | $1.908 \times 10^{-1}$ | $312.6 \pm 2.7$ | −80 to +55 | 20 |
| C IV | 1550.770 | $9.522 \times 10^{-2}$ | $177.9 \pm 7.8$ | −80 to +55 | 22 |
| N V | 1238.821 | $1.570 \times 10^{-1}$ | $43.5 \pm 5.5$[d] | −80 to +55 | 24 |
| N V | 1242.804 | $7.823 \times 10^{-2}$ | $28.7 \pm 4.5$[d] | −80 to +55 | 31 |
| O VI | 1031.926 | $1.325 \times 10^{-1}$ | $173.3 \pm 4.1$ | −71 to +76 | 20 |

[a] Vacuum wavelengths and oscillator strengths are taken from the compilation of Morton (1991).
[b] Equivalent widths are measured within the integration range listed in the adjacent column. The errors given are calculated by adding continuum placement errors and statistical errors in quadrature (see text).
[c] Signal-to-noise ratio measured in continuum next to each line.
[d] Additional error arises from large continuum placement uncertainty.

TABLE 5
TOTAL COLUMN DENSITIES FOR HIGH IONS

| Ion | $\lambda$ (Å) | log ($f\lambda$) | log $N_a$(cm$^{-2}$) −1σ | Best | +1σ | log $N$ (adopted) −1σ | Best | +1σ |
|---|---|---|---|---|---|---|---|---|
| Si IV | 1393.755 | 2.855 | 13.61 | 13.62 | 13.62 | 13.58 | 13.60[a] | 13.62 |
| Si IV | 1402.770 | 2.554 | 13.56 | 13.58 | 13.59 | | | |
| C IV | 1548.195 | 2.470 | 14.10 | 14.10 | 14.11 | 14.05 | 14.08[a] | 14.11 |
| C IV | 1550.770 | 2.169 | 14.03 | 14.05 | 14.07 | | | |
| N V | 1238.821 | 2.289 | 13.28 | 13.34 | 13.39 | 13.28 | 13.34[b] | 13.39 |
| N V | 1242.804 | 1.988 | 13.38 | 13.45 | 13.51 | | | |
| O VI | 1031.926 | 2.136 | 14.27 | 14.28 | 14.29 | 14.27 | 14.28 | 14.29 |

[a] Since the weaker lines of Si IV and C IV give smaller integrated column densities than the strong lines, we conclude that little saturated structure exists, and we have averaged the two estimates to obtain our adopted values of log $N$.
[b] We report the uncertain value of log $N$(N V) based on the strong N V absorption. The uncertain continuum makes it difficult to determine reliable errors.

TABLE 6
RESULTS OF COMPONENT FITTING TO HIGH ION PROFILES

| Ion | Component number[a] | $v_{LSR}$ (km s$^{-1}$) | $b$ (km s$^{-1}$) | $N$ (cm$^{-2}$) | $T_{max}$[b] ($10^5$ K) |
|---|---|---|---|---|---|
| Si IV | 1 | $-23.6 \pm 1.7$ | $26.3 \pm 0.9$ | $(2.42 \pm 0.12) \times 10^{13}$ | 12.4 |
| | 2 | $-36.3 \pm 0.7$ | $7.4 \pm 0.5$ | $(7.19 \pm 0.63) \times 10^{12}$ | 1.0 |
| | 3 | $-10.7 \pm 0.6$ | $5.2 \pm 0.6$ | $(4.51 \pm 0.58) \times 10^{12}$ | 0.57 |
| | 4 | $18.0 \pm 2.2$ | $12.9 \pm 2.0$ | $(3.78 \pm 0.67) \times 10^{12}$ | 3.7 |
| C IV | 1 | $-21.1 \pm 1.9$ | $34.8 \pm 1.0$ | $(1.03 \pm 0.04) \times 10^{14}$ | 9.2 |
| | 2 | $-36.1 \pm 1.2$ | $5.9 \pm 1.0$ | $(9.73 \pm 1.88) \times 10^{12}$ | 0.34 |
| | 3 | $-10.3 \pm 1.8$ | $3.6 \pm 1.7$ | $(2.97 \pm 1.63) \times 10^{12}$ | 0.20 |
| | 4 | $16.5 \pm 3.1$ | $10.6 \pm 3.3$ | $(6.32 \pm 2.28) \times 10^{12}$ | 1.4 |
| N V | 1 | $-15.6 \pm 2.6$ | $34.1 \pm 2.1$ | $(2.14 \pm 0.17) \times 10^{13}$ | 11.0 |
| O VI | 1 | $-2.3 \pm 0.1$ | $39.4 \pm 0.2$ | $(2.42 \pm 0.12) \times 10^{14}$ | 15.1 |

[a] Component 1 is the broad absorption seen in all four high ion profiles. Components 2 and 3 are the narrow features seen in Si IV and C IV, and Component 4 is the intermediate-width feature seen in Si IV and C IV.
[b] $T_{max}$ is a 1σ upper limit to the temperature determined from the ($b + 1\sigma$) value of the line width.

TABLE 7
COMPONENT-TO-COMPONENT COLUMN DENSITY RATIOS

| Component number | $v_{LSR}$(C IV)[a] (km s$^{-1}$) | $N$(C IV) / $N$(Si IV) | $N$(C IV) / $N$(N V)[b] | $N$(C IV) / $N$(O VI) |
|---|---|---|---|---|
| Total | … | 3.1 ± 0.2 | 5.7 ± 0.5 | 0.64 ± 0.03 |
| 1 | –21.1 | 4.3 ± 0.3 | 4.8 ± 0.4 | 0.54 ± 0.02 |
| 2 | –36.1 | 1.4 ± 0.3 | … | … |
| 3 | –10.3 | 0.7 ± 0.4 | … | … |
| 4 | 16.5 | 1.7 ± 0.7 | … | … |

[a] LSR velocity of component in C IV, which may be slightly different from the component velocity in the other ions (see Table 6).
[b] Any quantities calculated using $N$(N V) are subject to additional errors due to continuum placement uncertainties. Using the previous SS94 estimate of $N$(N V) reduces the overall $N$(C IV) / $N$(N V) ratio from 5.7 to 4.0.

TABLE 8
HIGH ION SCALE HEIGHTS AND DISK DENSITIES

| Ion | $I_p$[a] (eV) | $H$ (kpc) | $n_0$ ($10^{-9}$ cm$^{-3}$) | Ref.[b] | (log $N$)$_{mod}$ | (log $N$)$_{obs}$ | $n$(1.3 kpc) (cm$^{-3}$) |
|---|---|---|---|---|---|---|---|
| Si IV | 33.5 | 5.1 ± 0.7 | (2.3 ± 0.2) | 1 | 13.47 | 13.60 | 1.8 × 10$^{-9}$ |
| C IV | 47.9 | 4.4 ± 0.6 | (9.2 ± 0.8) | 1 | 14.06 | 14.08 | 6.8 × 10$^{-9}$ |
| N V | 77.5 | 3.9 ± 1.4 | (2.0 ± 0.5) | 1 | 13.39 | 13.34 | 1.4 × 10$^{-9}$ |
| O VI | 113.9 | 2.7 ± 0.4 | 20 | 2 | 14.36 | 14.28 | 1.2 × 10$^{-8}$ |

[a] $I_p$ = Creation ionization potential, from Moore (1970).
[b] References: (1) Savage et al. (1997); (2) Savage et al. (2000).

FIGURE CAPTIONS

Figure 1: Observed flux profiles for high ion lines in the spectrum of HD 116852 plotted against LSR velocity. The Si IV, C IV, and N V doublet profiles are STIS E140H observations, and the O VI line is a FUSE LiF1A observation. The labeled values for each panel on the ordinate correspond to the zero level for the panel above. Our fitted continua are shown with the short dashed lines; a long dashed line has been included at $v_{LSR}$ = 0 km s$^{-1}$ for reference. Other interstellar absorption lines falling in the regions shown are labeled in small font. The abrupt cut-off in C IV λ1550.770 is caused by the line falling at the upper end of a STIS echelle order.

Figure 2: Continuum-normalized flux profiles for the same lines as displayed in Figure 1. A short dashed line at a normalized flux value of unity indicates the assumed position of the continuum. Note how the narrow component structure prominent in Si IV becomes less evident as one moves toward the higher ions. The normalized O VI profile shown here has been treated for blending with a 15.6 mÅ line of HD λ1031.912.

Figure 3: Plots of apparent column density, $N_a(v)$ versus LSR velocity for the STIS high ion doublets and the FUSE O VI data. Note how the scales on the ordinate vary in factors of ten between each panel. The general agreement between the strong lines (open circles) and weak lines (filled circles) within the STIS data indicates that little saturated structure is present, and hence that the apparent column densities are good indicators of the true column densities along this sight line. The error bars on the STIS data (top three panels) were produced by measuring the r.m.s. dispersion in the data around the fitted continuum on either side of the line, interpolating the error for each data point within the line profile, and propagating the result through to column density space. The error bars on the FUSE data (bottom panel) were produced using the instrumental error array produced in the CALFUSE pipeline.

Figure 4: Comparison of normalized apparent column density plots between high ions. The Si IV and C IV profiles (top panel) display similar line shapes, albeit with differences in the shoulders at both negative and positive velocities. The central panel compares C IV λ1548.195 with O VI λ1031.926 – the O VI is clearly

centered at more positive velocities than the C IV, despite both profiles extending to similar negative velocities ($\approx$ 70 km s$^{-1}$). The bottom panel compares N V $\lambda$1238.821 with O VI $\lambda$1031.926. Each profile was arbitrarily normalized to its peak value, so that the line shapes could be easily compared.

Figure 5: Comparison of low (Mg II), intermediate (Al III), and high (Si IV) ions along the HD 116852 sight line. The Al III profile (obtained at a resolution of 11 km s$^{-1}$) was taken from archival GHRS data, whereas the Mg II and Si IV are from the STIS dataset discussed here. Each ion traces gas centered at different velocities, with the shoulder at –35 km s$^{-1}$ in Al III coinciding with Component 2 in Si IV.

Figure 6: Gaussian optical depth component fits to the normalized high ion absorption profiles. A simultaneous line fitting procedure was used to model the C IV and Si IV absorption, whereas an independent line fitting procedure was used to model the stronger lines in the the N V and O VI doublets. In each panel, the small circles represent the data points, the light lines display the individual components, and the heavy line represents the resulting model. In the case of the N V and O VI lines, only one component was needed, so the light and heavy lines are coincident. For the C IV and Si IV lines, two narrow components and an intermediate-width component are identified together with the broad absorption. The parameters of each model are given in Table 6. The two narrow components in C IV and Si IV at LSR velocities of –36 and –10 km s$^{-1}$ are likely produced in gas underneath the Norma-Centaurus and Sagittarius-Carina spiral arms, respectively.

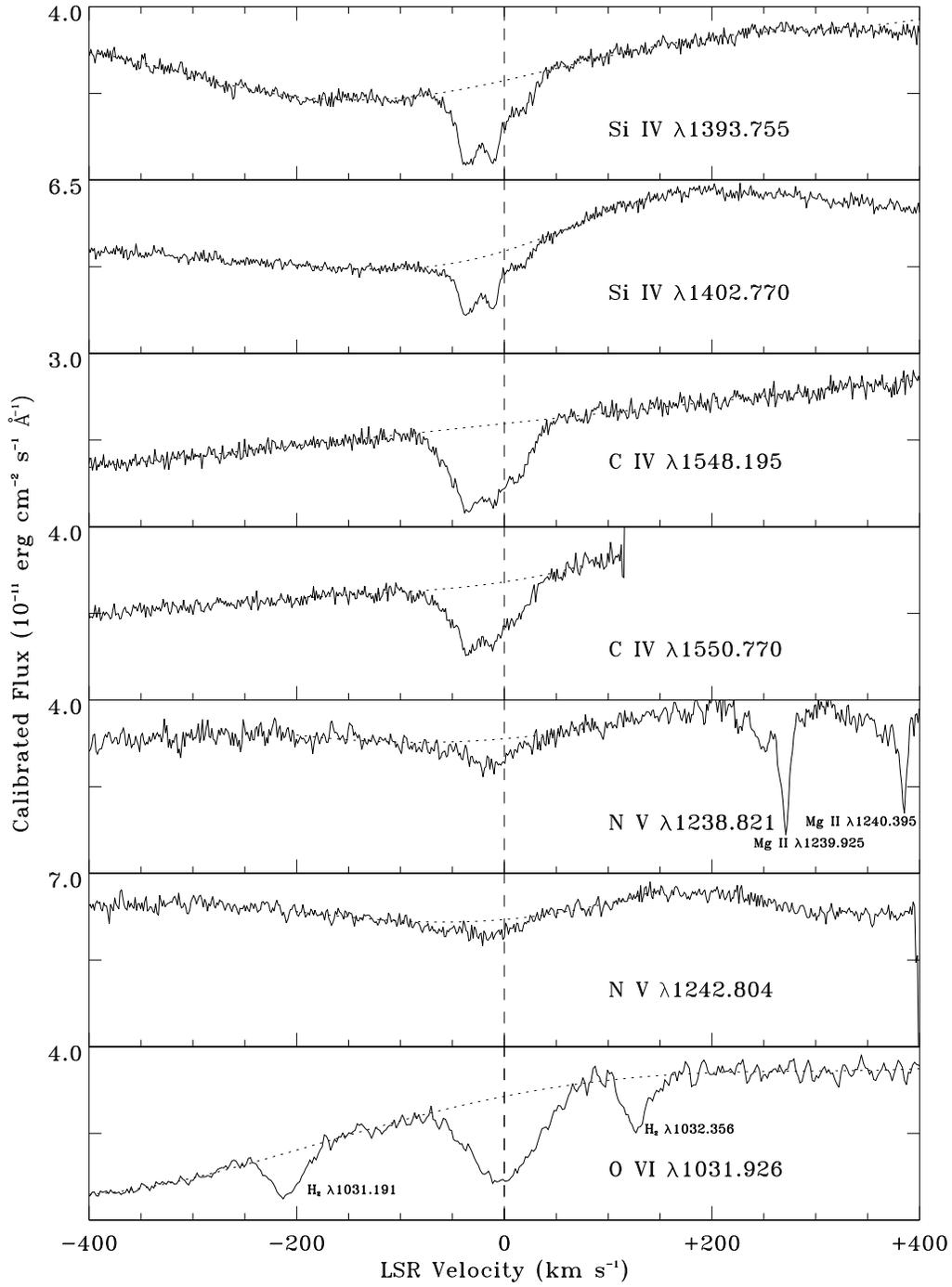

Fig. 1.— See text for caption

– 2 –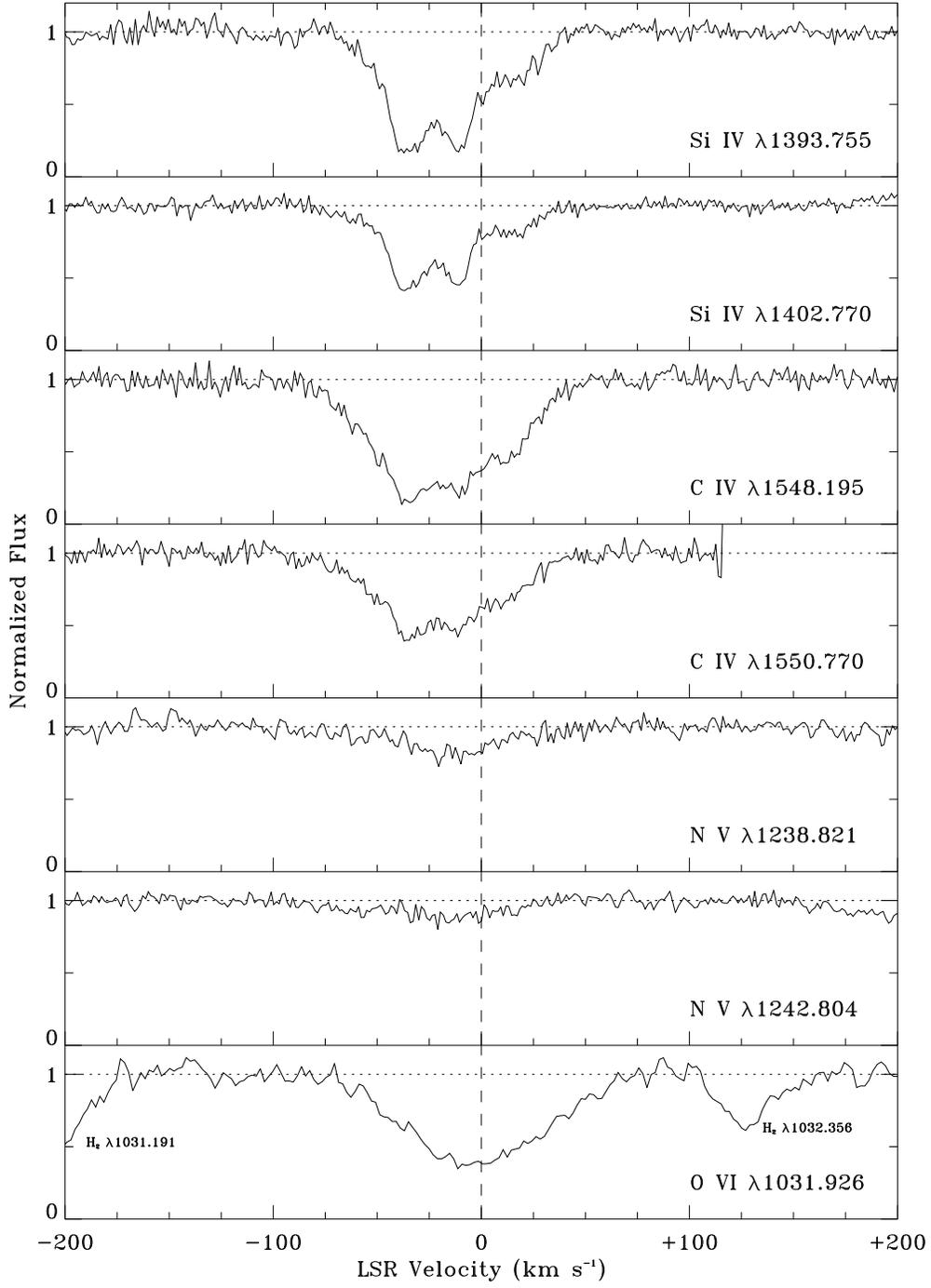

Fig. 2.— See text for caption

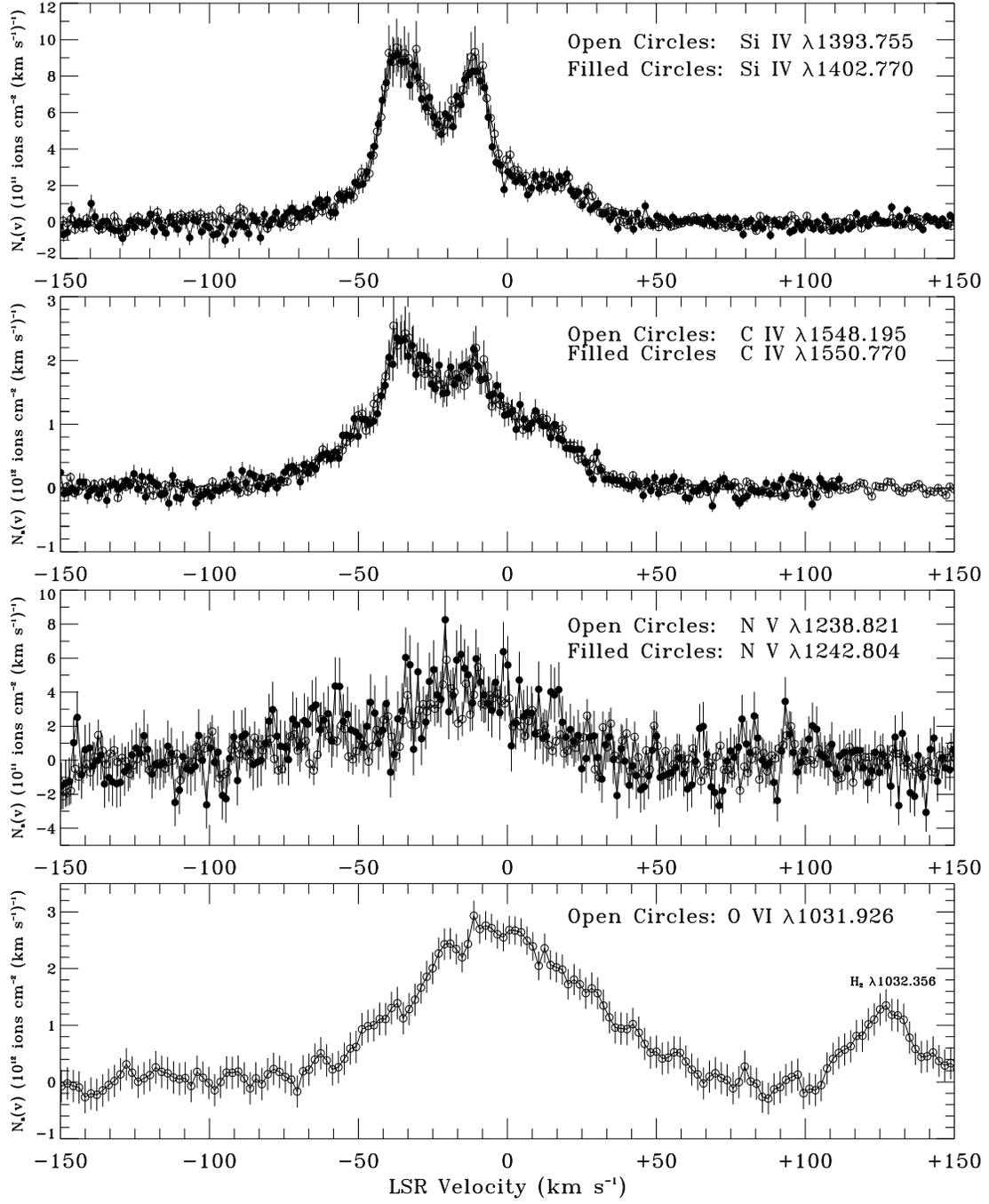

Fig. 3.— See text for caption

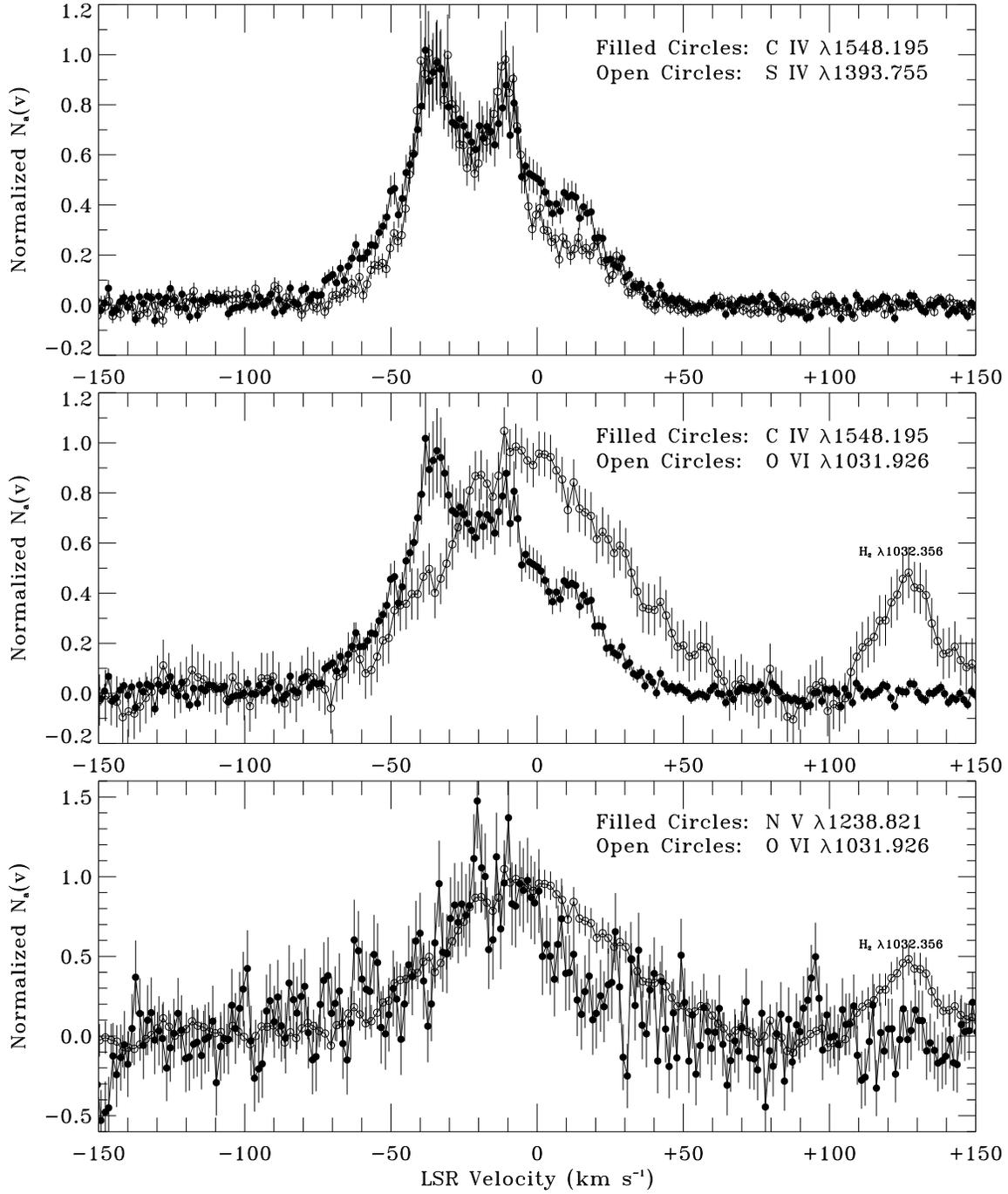

Fig. 4.— See text for caption

– 5 –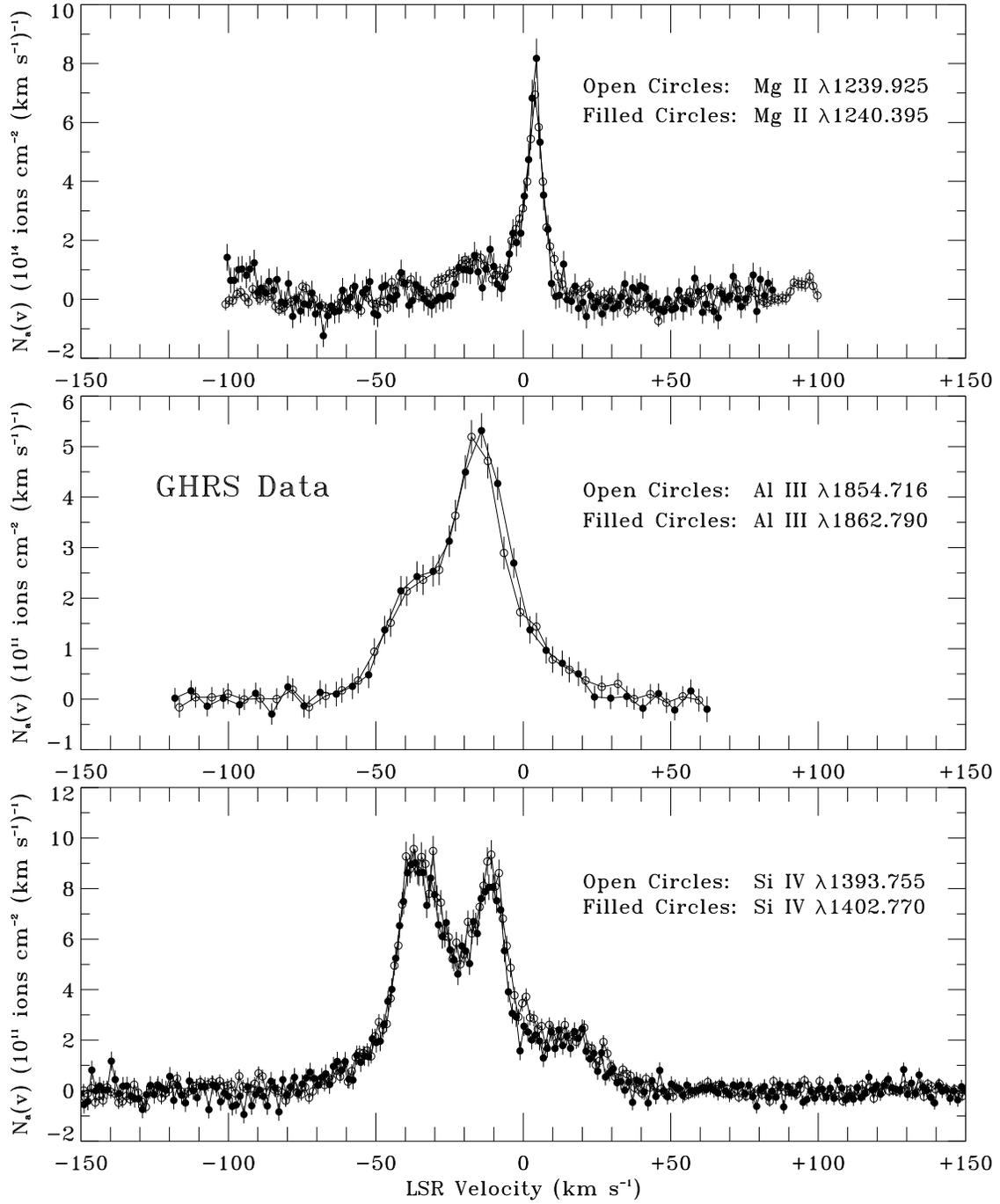

Fig. 5.— See text for caption

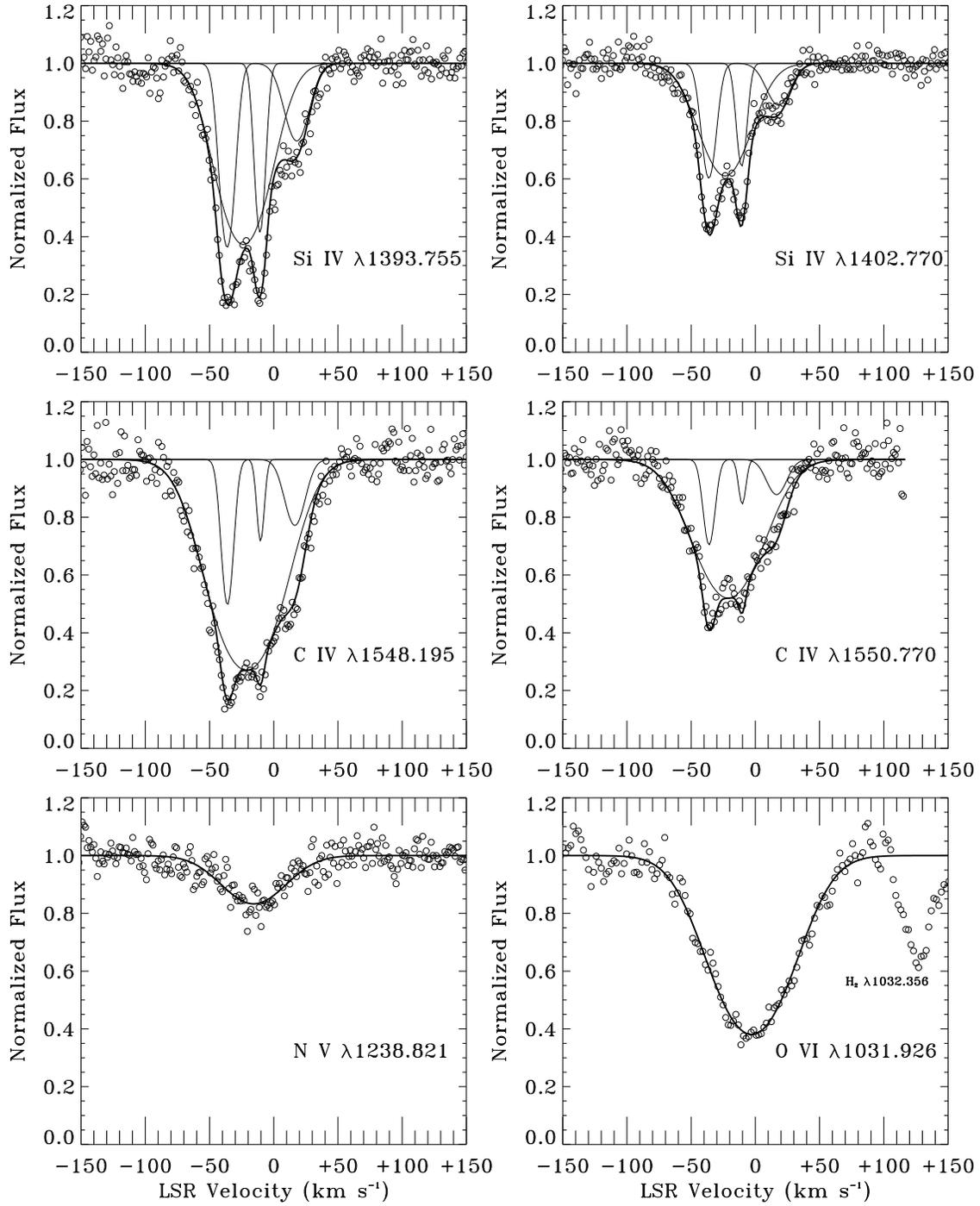

Fig. 6.— See text for caption